\documentclass[journal]{IEEEtran}
\IEEEoverridecommandlockouts

\usepackage{cite}
\usepackage{comment} 
\usepackage{amsmath,amssymb,amsfonts}
\usepackage{graphicx}
\usepackage{textcomp}
\usepackage{xcolor}
\usepackage{float}
\usepackage{graphicx}
\usepackage{setspace}
\usepackage{amssymb}
\usepackage[english]{babel}
\usepackage{acronym}
\usepackage{units}
\usepackage{multirow}
\usepackage{url}
\usepackage{sidecap}
\usepackage{todonotes}
\usepackage{algpseudocode}
\usepackage{algorithm}
\usepackage{subcaption}
\usepackage{tikz}

\DeclareRobustCommand{\bigO}{%
  \text{\usefont{OMS}{cmsy}{m}{n}O}%
}

\begin{document}

\title{{In-memory Associative Processors: Tutorial, Potential, and Challenges}
\author{ Mohammed E. Fouda, Hasan Erdem Yant{\i}r, Ahmed M. Eltawil, and Fadi Kurdahi
\vspace{-0.15in}}

\thanks{Mohammed Fouda and Fadi Kurdahi are with Center for Embedded \& Cyber-physical Systems, University of California-Irvine, Irvine, CA, USA 92697-2625. Email:foudam@uci.edu}
\thanks{H. Yant{\i}r is with TÜBİTAK – Informatics and Information Security Research Center, Gebze, 41470 Kocaeli, Turkey.}
\thanks{Ahmed Eltawil is with King Abdullah University of Science and Technology (KAUST), Thuwal, Saudia Arabia.}
}

\maketitle

\begin{abstract}
In-memory computing is an emerging computing paradigm that overcomes the limitations of exiting Von-Neumann computing architectures such as the memory-wall bottleneck. In such paradigm, {the computations are performed directly on the data stored in the memory, which highly reduces the memory-processor communications during computation. Hence, significant speedup and energy savings could be achieved especially with data-intensive applications.} Associative processors (APs) were proposed since the seventies and recently were revived thanks to the high-density memories. In this tutorial brief, we overview the functionalities and recent trends of APs in addition to the implementation of each {c}ontent-addressable memory with different technologies. The AP operations and runtime complexity are also summarized. We also explain and explore the possible applications that can benefit from APs. Finally, the AP limitations, challenges, and future directions are discussed. 
\end{abstract}

\section{Introduction}
Recently, most of the daily workloads are data-intensive, hence eliminating the need for memory-processor communications, i.e., memory-wall bottleneck, is crucial to enable orders of magnitudes speedup. In-memory computing (IMC) paradigm is the most promising candidate to achieve such speedup which is crucial for real-time on-the-edge processing. Generally speaking, IMC is referred to any kind of computations that are performed directly on the data inside the memory. In the literature, there are many ways to perform the IMC operations including crossbar arrays that inherently perform the matrix-vector multiplication in one clock cycle \cite{smagulova2021resistive} or by adding some peripheral circuits to the memory to support some operations such as logic operations and addition which is also referred to as near-memory computation. Although such techniques can achieve orders of magnitude speedup, they are dedicated to performing a single task only which is not preferred for general-purpose processors.

Associative processors rely on the associative memories to perform computations \cite{thurber1975associative,yau1977associative}, Hence, APs are considered as under the umbrella of IMC. The main advantage of associative processors is their ability to perform parallel processing, enabling word parallel, bit-serial operations that are needed for large data set manipulation. In addition, Associative processors offer the methodology to perform different types of computations, including conventional operations, inside the memory\cite{yantir2018hybrid,yantir2018two}. The importance of APs comes from the simplicity in performing single-instruction, multiple-data (SIMD) on associative memories. APs have been developed and explored since the seventies and have not been recognized due to limitations in developing large memories \cite{thurber1975associative}. During the last decade, APs have been revived due to two main reasons \cite{yavits2013computer,yavits2014resistive,guo2013ac}: 1) the emerging device technologies that enable high density and fast access time, and 2) the recent data-extensive applications such machine learning, visual computing, and natural language processing.

\begin{figure}[!t]
    \centering
    \includegraphics[width=0.8\columnwidth]{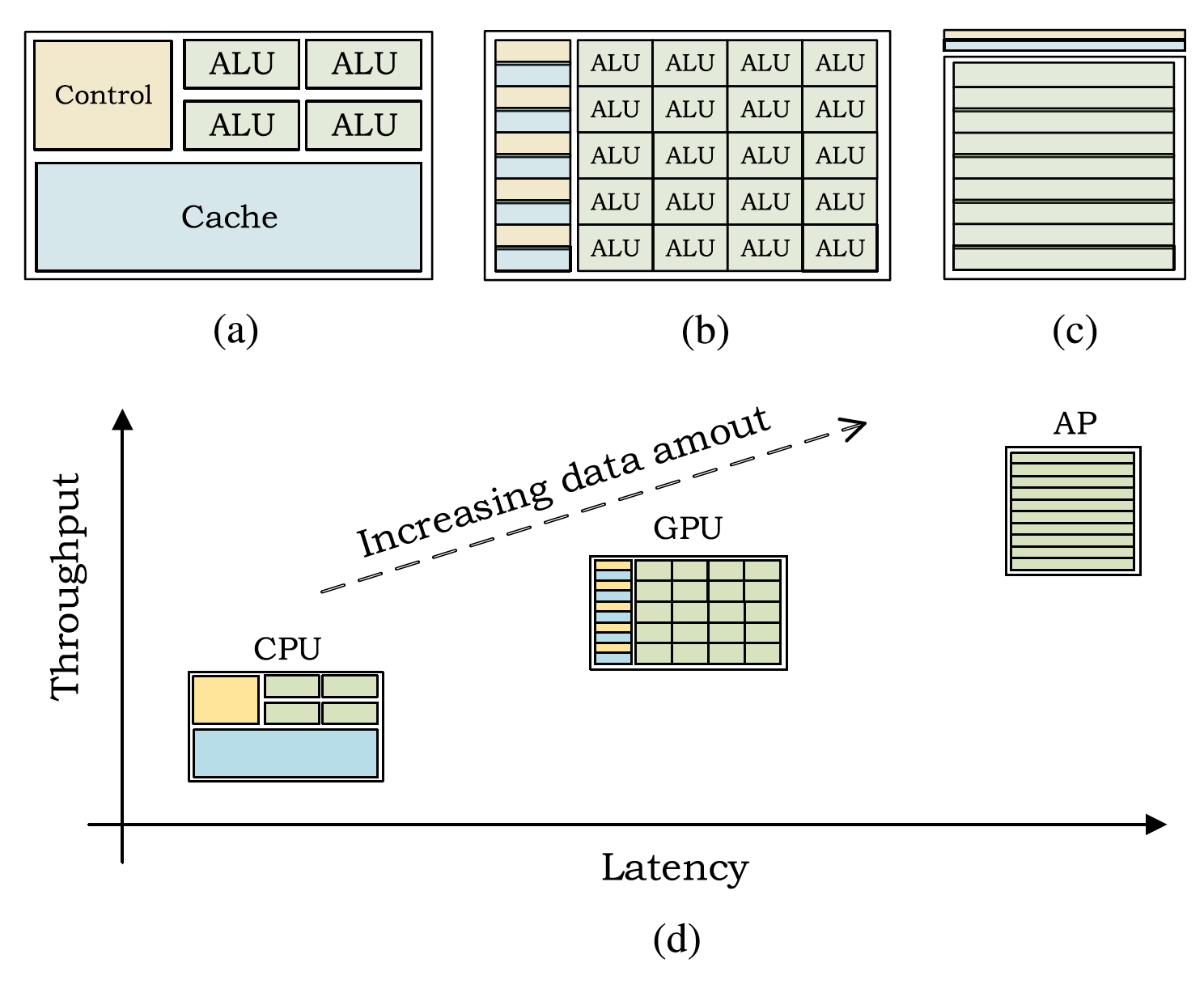}
        \vspace{-0.05in}
    \caption{An architectural comparison of (a) CPU, (b) GPU, and (c) AP architectures \cite{yantir2021imca}.}
    \label{fig:cpu_gpu_ap}
    \vspace{-0.15in}
\end{figure}

Recently, associative processors have shown efficiency in many applications alongside traditional computations, such as machine learning and deep learning acceleration \cite{yantir2021imca} and showing superior performance compared to CPUs and GPUs. To further enhance the energy and delay efficiency of associative processors, other computing techniques such as approximate computing can be run on top regular operation on the associative processor \cite{yantir2017approximate}.

In terms of performance, APs shine in the presence of large amounts of data outperforming all traditional processors relying on SIMD instruction set. The optimal applications for APs are the ones that have inherent SIMD computational pattern such as matrix multiplication \cite{yavits2014sparse,neggaz2018rapid}, Fast Fourier transform (FFT) \cite{yantir2019ultra}, DNA sequence alignment \cite{kaplan2017resistive}, multi-valued arithmetic \cite{hout2021memory}, Speech-to-speech translation \cite{higuchi1994ixm2}, etc. In contrast with traditional CPUs and GPUs that are designed to minimize the latency, the performance of APs is optimized to maximize the throughput due to their single memory structure and simplicity of the controller \cite{yantir2018hybrid}. Fig. \ref{fig:cpu_gpu_ap} depicts an architectural comparison of the these processors. In this tutorial, we will discuss the current trends in designing APs and functionality in addition to hardware challenges and potential applications. We will also survey recent works on devices that could enhance AP energy and delay performance.  

The tutorial is organized as follows: Section \ref{Sec:AP_Arch} discusses the AP architecture, functionality, advantages and possible content-addressable memory implementations. The AP applications are introduced in Section \ref{Sec:AP_apps}. Finally, the AP limitations, challenges and Future directions are discussed in Sections \ref{Sec:limitations} and \ref{Sec:future_Dir}, respectively.

\section{AP Architecture and Organization}
\label{Sec:AP_Arch}
\begin{figure}
    \centering
    \includegraphics[width=\columnwidth]{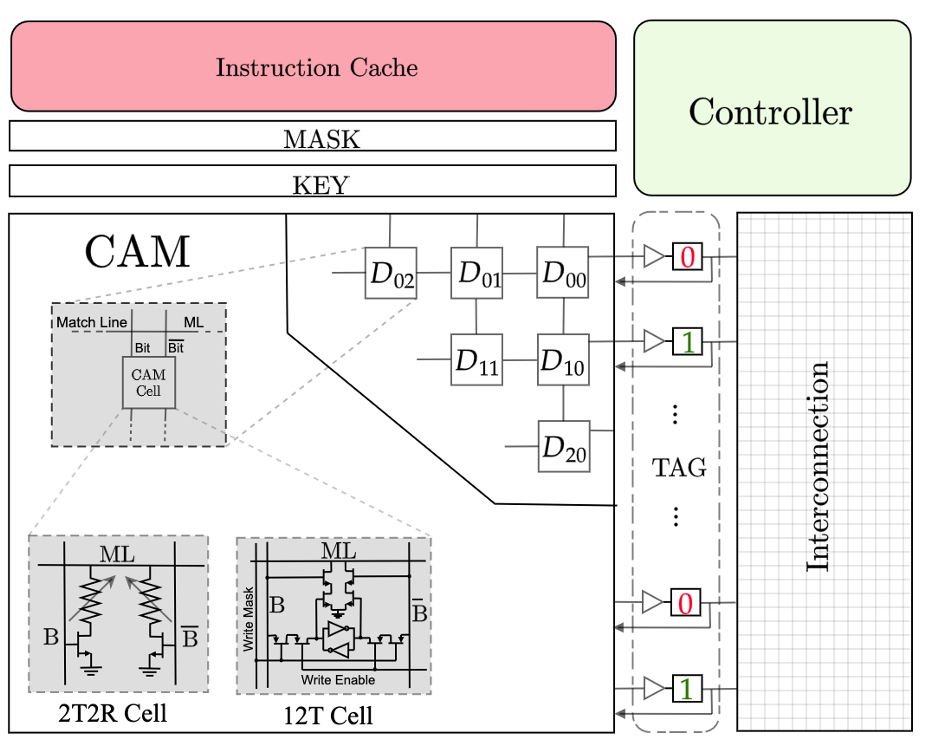}
    \caption{General Associative Processor Architecture.}
    \label{fig:AP_arch}
    \vspace{-0.1in}
\end{figure}

Presented in Fig. \ref{fig:AP_arch}, the general architecture of an AP originated in the seventies and eighties \cite{foster1976content,scherson1989reconfigurable,yavits2013computer}. AP comprises content addressable memory (CAM), controller, instruction cache, interconnection circuit, and specific registers including mask, key, and tag registers. The CAM is mainly used to store the data to be searched/ processed. The CAM is structured as a 2-D matrix where each cell can be accessed by activating the corresponding bit lines. 
The controller generates the required mask and key values for each corresponding instruction. The key register is used to present the value that is written or compared against. The mask register indicates which bits are activated during the compare and write cycles. The rows matched by the compare operation are marked in the tag field where the rows tagged with logic-1 means that the corresponding CAM row has been matched with the given key and mask value.  For example, if we send key 100 and mask 101 to the CAM, the tag bits of the corresponding rows, whose first and third bits are logic-0 and logic-1 respectively, become logic-1. Operations on AP consist of consecutive compare and write phases. During the compare phase, the matched rows are selected, and in the write phase, the corresponding masked key values are written into tagged CAM words.

The interconnection matrix is a basic switching matrix that allows the AP rows to communicate in parallel either bitwise or wordwise. Recent works consider memristive crossbar arrays as switching matrix \cite{yantir2018efficient} instead of the conventional  CMOS-based reduction tree, as in \cite{yavits2014resistive,guo2013ac,yavits2013computer}, which improve both AP utilization and the area leading to significant improvement in the performance. In addition, the memristive switching matrix provides higher density and better functionality, unlike the reduction trees that do not allow the intercommunication between the AP rows complicating certain operations including array multiplication \cite{yantir2018efficient}. 

Generally, the CAM can be used as associative memory which is very handy for many applications such as decision trees \cite{pedretti2021tree} or speech translation \cite{higuchi1994ixm2}, where compare cycles only are needed. But, to perform any computation, compare and write cycles are required.

\textit{\textbf{Compare Cycle}} is divided into two phases, the first phase is a \textit{precharge} phase where the matchline is precharged to the supply voltage. And the second cycle is an \textit{Evaluate} cycle where the searched inputs are applied to the columns creating a low resistive path to the ground when the stored data do not match the searched data and vice versa. The matchline is discharged when there is at least one mismatched cell leaving a considerable voltage gap to the sensing circuits to differentiate the match and mismatch cases. The analysis of the optimal evaluate latency and energy is discussed in detail in \cite{bahloul2017design}. It is worth highlighting that another way to sense the matchline status is through resistive sensing in which the precharge cycle is not needed offering faster evaluations. But, the main drawback is the lower sensing margin compared to capacitive sensing \cite{rakka2020design}.  

\textit{\textbf{Write Cycle}} is used to update the CAM cell based on a lookup table that is constructed to perform a certain function. In the case of in-place operation, the operand bits are rewritten based on the lookup table as will be discussed in the next subsection.

\begin{table}[!t]
\centering
\caption{Promising technologies for CAM implementations. Reported numbers for the best devices in the literature \cite{wang2020resistive}.}
\label{tab:CAMCell}
\resizebox{0.5\textwidth}{!}
{%
\begin{tabular}{lccccc}
\cline{2-6}
\multicolumn{1}{l}{\multirow{2}{*}{}}                                          & \multicolumn{4}{c}{2T2R}      & \multicolumn{1}{c}{\multirow{2}{*}{SRAM}}                        \\ \cline{2-5}
\multicolumn{1}{l}{}                & \multicolumn{1}{c|}{Redox}         & \multicolumn{1}{c|}{PCM}   & \multicolumn{1}{c|}{MTJ}                                          & \multicolumn{1}{c}{FeD}                                             & \multicolumn{1}{c}{}                                          \\ \hline
\multicolumn{1}{l|}{Max Endurance (Cycles)}                                   & \multicolumn{1}{c|}{$10^{12}$}    & \multicolumn{1}{c|}{$10^{11}$}     & \multicolumn{1}{c|}{$10^{12}$}     & \multicolumn{1}{c|}{$4\times 10^{6}$}                                           & \multicolumn{1}{c}{$>10^{16}$}      \\ \hline
\multicolumn{1}{l|}{Min. Switching Energy (fJ)}     & \multicolumn{1}{c|}{$115$}      & \multicolumn{1}{c|}{$1000$}       & \multicolumn{1}{c|}{$10$}         & \multicolumn{1}{c|}{$100$}       & \multicolumn{1}{c}{$1-100$}        \\ \hline
\multicolumn{1}{l|}{Max Switching Speed (ns)}    & \multicolumn{1}{c|}{$0.085$}      & \multicolumn{1}{c|}{$0.7$}        & \multicolumn{1}{c|}{$0.2$}         & \multicolumn{1}{c|}{$10$}        & \multicolumn{1}{c}{$0.1-0.25$}       \\ \hline
\multicolumn{1}{l|}{\begin{tabular}[c]{@{}l@{}}Min. CAM cell Area${}^\dagger$\\ ($\mu m^2$) @ min feature dim.\end{tabular}} & \multicolumn{1}{c|}{\begin{tabular}[c]{@{}c@{}}$0.0014$\\ $@2nm$\end{tabular}} & \multicolumn{1}{c|}{\begin{tabular}[c]{@{}c@{}}$0.0014$\\ $@5nm$\end{tabular}} & \multicolumn{1}{c|}{\begin{tabular}[c]{@{}c@{}}$0.0014$\\ $@10nm$\end{tabular}} & \multicolumn{1}{c|}{\begin{tabular}[c]{@{}c@{}}$0.0032$\\ $@20nm$\end{tabular}} & \multicolumn{1}{c}{\begin{tabular}[c]{@{}c@{}}$0.042$\\ $@5nm$\cite{yeap20195nm}\end{tabular}} \\ \hline
\multicolumn{1}{l|}{Retention Time (Years)}                              & \multicolumn{1}{c|}{$>1,000$}   & \multicolumn{1}{c|}{$>1,000$}    & \multicolumn{1}{c|}{$>10$}      & \multicolumn{1}{c|}{$>100$}        & \multicolumn{1}{c}{-}                                    \\ \hline
\multicolumn{6}{l}{\begin{tabular}[c]{@{}l@{}}${}^\dagger$ Transistor area is estimated to be $8  \times 14\lambda^2$  and $\lambda=2.5nm$ for $5nm$ technology\\ node. The RD area is estimated to be 4 times the minimum dim. squared. The \\maximum area  between the transistors and RD is taken.\end{tabular}} \end{tabular}%
}
\vspace{-0.25in}
\end{table}

\begin{figure*}[!t]
    \centering
    \includegraphics[width=0.8\textwidth,height=0.6\columnwidth]{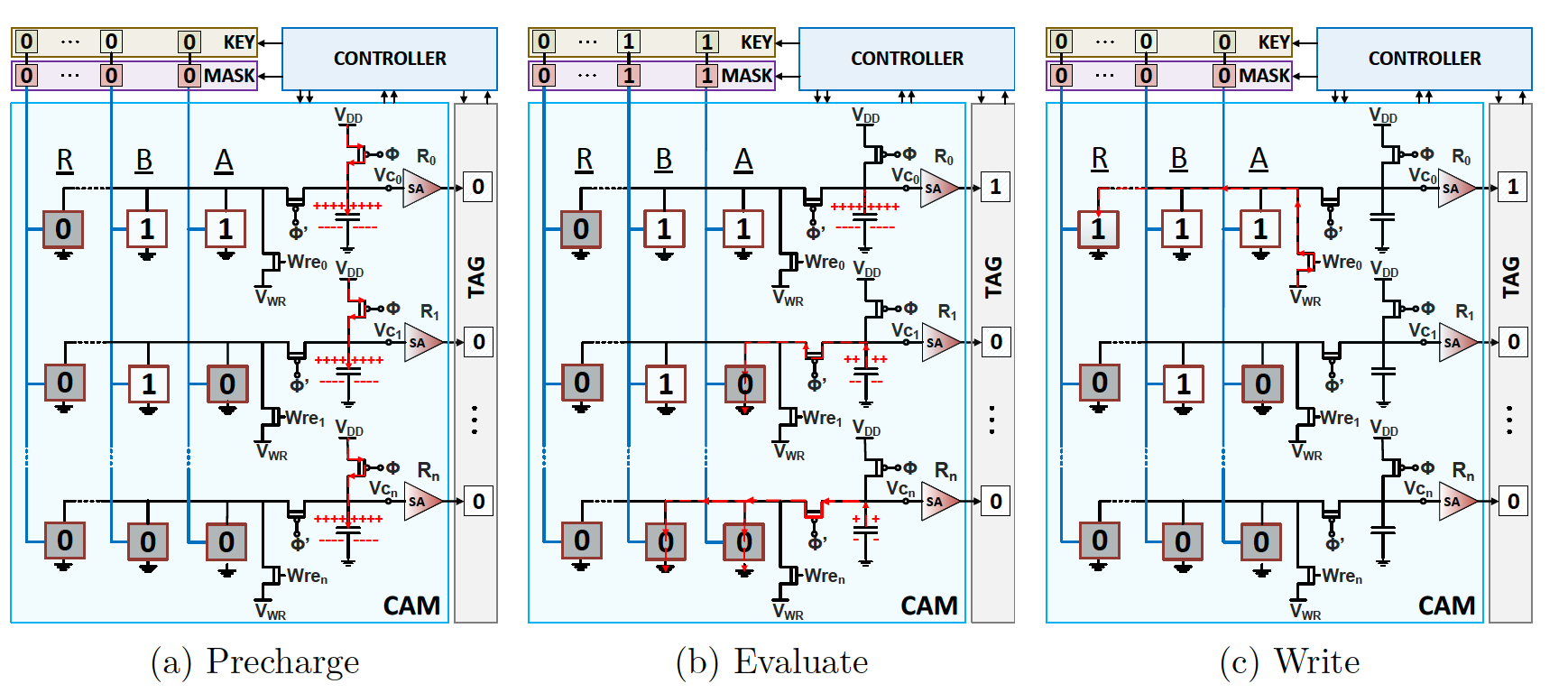}
    \caption{Execution of an AND operation on AP.}
    \label{fig:AND_AP_operation}
    \vspace{-0.15in}
\end{figure*}

\subsection{Illustrative Example on AP}
{Utilizing consecutive compare and write cycles, any SIMD function, that is performed on a sequential processor, can also be performed on APs in parallel. To do any computation on AP, a lookup table (LUT) needs to be constructed first. An example of two LUTs for an AND operation and an in-place addition is depicted in Table. \ref{tab:LUTexample}. It is worth mentioning that some operations can be performed in-place such as addition or subtraction, while most of the operations are usually done out-of-place such as logic operations and multiplication. More examples on logic and arithmetic operations can be found in \cite{yantir2018efficient}. The AP goes through this LUT in a certain order to avoid having any cycles that could overwrite the outcome of earlier passes. Hence, the pass order is crucial. In \cite{hout2021memory},  an automated way was proposed to construct the LUT for any operation where a state diagram is built and has to be uni-directional with no cycles. In the case of having cycles in the state diagram, the cycles are broken by redirecting the backward edges forward to a state having the same output \cite{hout2021memory}. Once, the LUT is constructed, the corresponding masks are generated to perform the compare and write cycles.}  

{As an example of the functionality of the AP is discussed on how a parallel AND ($R = A \& B$) operation is performed. As shown in Table \ref{tab:AND}, the AND operation has a single pass which would require a single compare cycle to search for "11" and a single write cycle to update the output cell. Figure \ref{fig:AND_AP_operation} shows an AP with $3 \times 3$ CAM where the first and second columns correspond to the $A$ and $B$, respectively, and the last column is R where its all cells are initially set to "0". The LUT operation in the CAM can be simply performed by looking for "11" in the $A$ and $B$ columns of the CAM and writing logic-1 to the result column of the matched rows as indicated by the AND. Since the other combinations of $A$ and $B$ do not affect the R which is already set to "0".} 

{During the compare operation, all rows of the CAM are pre-charged, as shown in Fig. \ref{fig:AND_AP_operation}a followed by the evaluate operation. In the evaluate phase, the $A$ and $B$ columns are searched for "11" where the Mask bits corresponding to $A$ and $B$ are activated as well (i.e., set to "11") where $R$ mask is set to "0".
The LUT of the AND operation is applied to the CAM where key bits corresponding to columns $A$ and $B$ are set to "1" as shown in Fig. \ref{fig:AND_AP_operation}b. The second and the third rows leak the charge stored in the capacitor while the first row only matches with the given key and mask values. The sense amplifier is used to sense the matchline of each row by comparing the capacitor's voltage to a threshold voltage and stores "1" or "0" in the TAG bit for the match or the mismatch, respectively. Then, in the write cycle, a write signal is generated to set the $R$ cell to "1" for the matched rows as shown in Fig. \ref{fig:AND_AP_operation}c. The same procedure is repeated for the other passes on the same columns or other columns as well.}   

\begin{table}[t]
\caption{Examples of the AP lookup tables. }
\begin{subtable}[h]{0.4\columnwidth}
 \caption{LUT for AND}
 \centering
\resizebox{\columnwidth}{!}{%
\begin{tabular}{cc|cl}
\hline
$B$ & $A$ & $R$ & Comment   \\ \hline \hline
$0$ & $0$ & $0$ & $NC$      \\ \hline
$0$ & $1$ & $0$ & $NC$      \\ \hline
$1$ & $0$ & $0$ & $NC$      \\ \hline
$1$ & $1$ & $1$ & $1stPass$ \\ \hline
\end{tabular}

\label{tab:AND}
}
\end{subtable}
\hfill
\begin{subtable}[h]{0.6\columnwidth}
\caption{LUT for In-place Addition}
\centering
\resizebox{0.9\columnwidth}{!}{%
\begin{tabular}{ccc|ccc}
\hline
$C_r$ & $B$ & $A$ & $C_r$ & $B$ & Comment   \\ \hline \hline
$0$   & $0$ & $0$ & $0$   & $0$ & $NC$      \\ \hline
$0$   & $0$ & $1$ & $0$   & $1$ & $2ndPass$ \\ \hline
$0$   & $1$ & $0$ & $0$   & $1$ & $NC$      \\ \hline
$0$   & $1$ & $1$ & $1$   & $0$ & $1stPass$ \\ \hline
$1$   & $0$ & $0$ & $0$   & $1$ & $3rdPass$ \\ \hline
$1$   & $0$ & $1$ & $1$   & $0$ & $NC$      \\ \hline
$1$   & $1$ & $0$ & $1$   & $0$ & $1stPass$ \\ \hline
$1$   & $1$ & $1$ & $1$   & $1$ & $NC$      \\ \hline
\end{tabular}
}
\label{tab:Add}
\end{subtable}
\label{tab:LUTexample}
\vspace{-0.15in}
\end{table}

\subsection{Possible CAM Implementation}
\label{Sec:CAM_imp}

Since the CAM is the main building block, the denser the CAM cell the better. Hence, different technologies were considered to realize the CAM including pure digital implementations such as Flipflops \cite{kokosinski2002fpga}. Recent advances in emerging device technologies in resistive switching devices opened the road for high dense memories and associative memories \cite{yavits2014resistive}. Such resistive devices (RDs) \cite{wang2020resistive} include Redox devices such as ReRAM, phase change memory, magneto resistive such as magnetic tunneling junction, and Ferro-electric devices such as FeFETs\cite{yin2020fecam}.  Such devices can be used with a transistor in a 2T2R structure to construct a ternary CAM cell as shown in the subplot of Fig. \ref{fig:AP_arch}. The data \{"0", "1"\} are encoded into two RDs; each is programmed to a high or low resistance state. Hence, the left and right devices are programmed to high and low resistances for "0", respectively, and the opposite for "1". Each RD is accessed through a transistor with a bitline signal (B). While the searched data are encoded as follows "0" is mapped to "10" and "1" is mapped to "01" while the don't care ("X") is mapped to "00".  The searched data are sent through the bitline signals to access the cell. In case of a mismatch, RD is in a low resistance state and its transistor is turned on. Thus, a low resistive path is created pulling down the matchline to the ground. On the other hand, in case of a match case, either the RD or transistor or both is kept in an off state preventing any leaking from the matchline. In case of a match, the TAG cell is programmed to 1 and a write signal is sent to update the CAM cell.

Table \ref{tab:CAMCell} shows the best RDs in the literature in terms of each metric \cite{wang2020resistive}. Due to the maturity of the 5 nm technology node nowadays, we report all the numbers scaled to this node. It is worth mentioning that we focus on switching (i.e., write) energy, speed, cell area and the endurance cycles since they highly affect the total performance of the AP. Other metrics such as read energy or latency are much smaller than the switching energy or latency. Clearly from the table, each RD technology has some advantages over the others. According to that table, Redox (i.e., ReRAM) is the most promising technology for APs. For instance, a 64-kb fully integrated ReRAM-CMOS CAM with TSMC 40-nm ReRAM technology is introduced in \cite{li2021sapiens}. Other fabricated ReRAM macros is summarized in Table III in \cite{smagulova2021resistive}. 

Before the rise of resistive technologies, SRAM has been considered the way to implement dense CAM arrays.
Several SRAM topologies can be used to realize a CAM \cite{stormon1992general,pagiamtzis2006content,grosspietsch1992associative}. The operation of the SRAM is the same as regular SRAM-based memories with 4 extra transistors to pull down the matchline under mismatch scenario and to program the cell when a match occurs. An example of a 12T SRAM-based CAM cell is shown in Fig. \ref{fig:AP_arch} \cite{stormon1992general}. The same aforementioned data encoding is used for SRAM as well. As shown in Table \ref{tab:CAMCell}, SRAM-based cell is excellent in write energy, latency, and endurance except the cell area which is around $30 \times$ the area of 2T2R.

It is worth mentioning that other technologies such as DRAM can be used to realize CAM. But, they are not as efficient as resistive technologies or SRAM in terms of density or energy, or speed. In this work, the resistive device-based AP is referred to as RAP and the SRAM-based AP is referred to as SAP.

\subsection{AP Advantages}
\subsubsection{Vector Operations}
As shown in the example, the computations are done bitwise and the same operation is done on the entire column. Hence, the runtime complexity is a function of the bit precision only and is equivalent to the number of passes. Table \ref{tab:complexity} depicts the runtime, area usage per CAM row (in terms of bits), and algorithmic complexities of different arithmetic and logical operations. The runtime of a vectorial operation on $m$-bit and $n$ numbers depends only on the number of bits ($m$), not on the number of vectors ($n$). Hence, the larger the CAM, the higher throughput.

\subsubsection{Runtime-bit Precision Configuration}
As previously discussed, the operations are performed bitwise which would give a major advantage to AP where the computation precision can be changed during the runtime. Such an advantage does not exist in regular processors. APs can be tuned to certain precision to maximize throughput or minimize power or latency. An example of a runtime precision tuning is discussed in \cite{yantir2021imca} where the computational performance in terms of operations per sec could increase 4x when half-precision is used. This is also particularly relevant in applications that naturally tolerate imprecise computation levels such as image and video processing \cite{1613133}\cite{5090795}. In \cite{yantir2017approximate} it was also demonstrated that  AP-based approximate resistive in-memory computing improves the AP's energy efficiency (up to 80x) and performance (up to 20x) on a variety of benchmarks from different domains when output quality degradation is limited to 10\%.

\begin{table}[!t]
\caption{Running time and area evaluation of primitive AP operations/instructions.}
\resizebox{\columnwidth}{!}
{%
\begin{tabular}{rccc}
\textbf{Function}      & \textbf{Runtime}     & \textbf{Area/Row}    & \textbf{Complexity}              \\ \hline
NOT                    & $2m$                 & $2m$                 & $\bigO(m)$                       \\
AND                    & $2m$                 & $3m$                 & $\bigO(m)$                       \\
OR                     & $6m$                 & $3m$                 & $\bigO(m)$                       \\
XOR                    & $6m$                 & $3m$                 & $\bigO(m)$                       \\ \hline
Addition (IP, S/U)     & $10m$                & $2m+1$               & $\bigO(m)$                       \\
Addition (OOP, S/U)    & $11m$                & $3m+1$               & $\bigO(m)$                       \\ \hline
Subtraction (IP, S/U)  & $10m$                & $2m+1$               & $\bigO(m)$                       \\
Subtraction (OOP, S/U) & $11m$                & $3m+1$               & $\bigO(m)$                       \\ \hline
2's Complement (OOP)   & $6m$                 & $2m+1$               & $\bigO(m)$                       \\
Absolute Value (OOP)   & $8m$                 & $2m+1$               & $\bigO(m)$                       \\ \hline
Multiplication (U)     & $10m^2$              & $4m$                 & $\bigO(m^2)$                     \\
Multiplication (S)     & $10m^2+4m-14$        & $8m+4$               & $\bigO(m^2)$                     \\
MAC (U)                & $10m^2+10m$          & $4m$                 & $\bigO(m^2)$                     \\
MAC (S)                & $10m^2+14m-14$ & $8m+4$ & {$\bigO(m^2)$}\\ \hline
\multicolumn{4}{l}{*IP: in-place, OOP: out-of-place, S: Signed, U:unsigned, m:bitwidth.}
\end{tabular}%
}
\vspace{-0.15in}
\label{tab:complexity}
\end{table}

\subsection{2-D Associative Processor}
To further improve the functionality of the AP to achieve lower latency and higher throughput, the associative memory can be extended in two dimensions where an extra vertical key is added to do vector computations on both vertical and horizontal dimensions \cite{yantir2018two}. Such extension is very useful in applications that require matrix multiplication such as image filtering and 2-D transformations. Table \ref{tab:task_complexity} shows theoretical bounds of runtime complexity of various kernels running on 2-D AP and the best 1-D AP. In some cases, the 2-D AP achieves the same complexity as the 1-D AP while in other cases, a significant speedup
can be attained due to the concurrent computation on the data and data movement savings which requires additional interconnections in 1-D AP. Despite, the increase in the circuits' complexity and the need for a more complex controller for the 2-D AP leading to an increase in the power and area, 2-D AP offers up to 80x performance improvement in the throughput and energy savings compared to 1-D AP for various applications including FFT and filtering \cite{yantir2018two}. 

\begin{table}[]
\caption{Theoretical complexity of various kernels.}
\resizebox{\columnwidth}{!}{%
\begin{tabular}{lccc}
\hline
Kernel                                      & Sequential           & 1D AP \cite{yavits2014resistive,ipek2014resistive,guo2011resistive} & 2-D AP \cite{yantir2018two} \\ \hline
Vector Add, Multiply                        & $\bigO(n)$           & $\bigO(1)$                             & $\bigO(1)$                         \\ \hline
Vector Dot Product                          & $\bigO(n)$           & $\bigO(\log_2 n)$                       & $\bigO(\log_2 n)$                   \\ \hline
Vector-Matrix Multiply                      & $\bigO(n^2)$         & $\bigO(n \log_2 n)$                     & $\bigO(\log_2 n)$                   \\ \hline
Matrix-Matrix Multiply                      & $\bigO(n^3)$         & $\bigO(n^2 \log_2 n)$                   & $\bigO(n \log_2 n)$                 \\ \hline
Histogram ($k$ bins and $n$ values)         & $\bigO(kn)$          & $\bigO(k)$                             & $\bigO(k)$                         \\ \hline
Frequency ($k$ bins in $n$ values)          & $\bigO(kn)$          & $\bigO(k)$                             & $\bigO(k)$                         \\ \hline
Set membership                              & $\bigO(n)$           & $\bigO(1)$                             & $\bigO(1)$                         \\ \hline
Set intersection or union                   & $\bigO(n^2)$         & $\bigO(n)$                             & $\bigO(n)$                         \\ \hline
1D FFT, IFFT, FWHT, Conv. ($n$)             & $\bigO(n \log_2 n)$   & $\bigO(\log_2 n)$                       & $\bigO(\log_4 n)$                   \\ \hline
1D Filter ($n$ data, $m$ filter)                & $\bigO(n)$           & $\bigO(1)$                             & $\bigO(1)$                         \\ \hline
2-D FFT ($n\times n$)                        & $\bigO(n^2 \log_2 n)$ & $\bigO(n \log_2 n)$                     & $\bigO(n \log_4 n)$                 \\ \hline
2-D Stencil ($n\times n$, 5 point)           & $\bigO(n^2)$         & $\bigO(n)$                             & $\bigO(1)$                         \\ \hline
2-D Filter ($n\times n$, $m\times m$ filter) & $\bigO(n^2)$         & $\bigO(n)$                             & $\bigO(1)$                         \\ \hline
\multicolumn{4}{l}{$m$ is the bitwidth and is constant.}
\end{tabular}%
}
\label{tab:task_complexity}
\vspace{-0.15in}
\end{table}

\section{APs benchmarks and Applications}
\label{Sec:AP_apps}

\subsection{AP Performance against Existing Accelerators}

\begin{figure}[!t]
     \centering
         \centering
         \includegraphics[width=0.9\columnwidth]{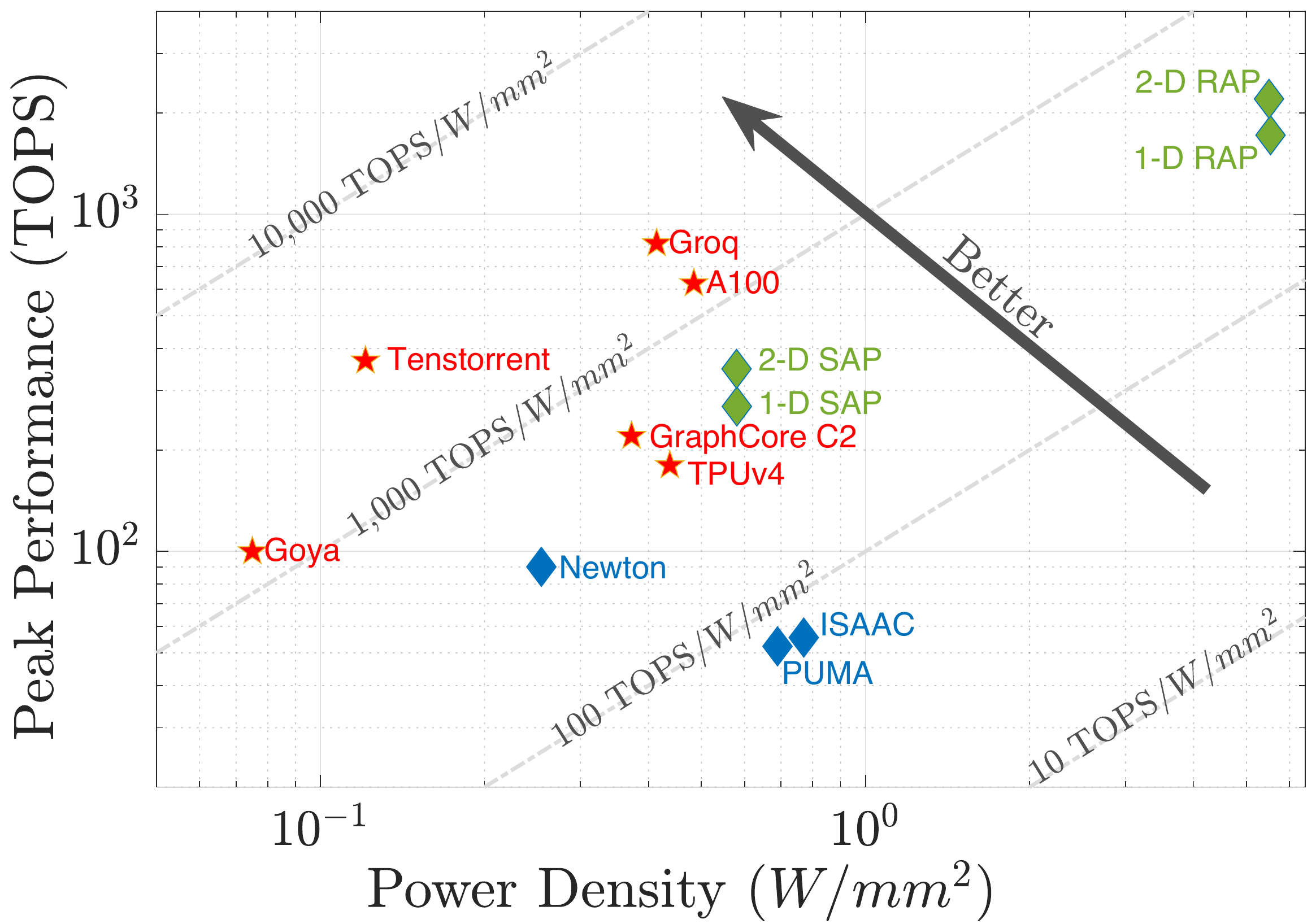}
\caption{Peak performance of the associative processor against state-of-the-art commercial accelerators and some resistive accelerators.}
\label{fig:performance}
\vspace{-0.15in}
\end{figure}

Figure \ref{fig:performance} shows the peak performance in Tera operations per second (TOPS) versus the power density ($W/mm^2$). It is worth mentioning that the used numbers for the RAP and SAP are based on the currently available 
technologies (i.e., 16 nm node) \cite{yantir2018two} and not based on the reported numbers in Table \ref{tab:CAMCell}. Although RAP is outperforming SAP in the peak performance by $6.3\times$, the RAP is having one order of magnitude higher than SAP in the power density. While keeping the similar numbers for the peak performance per power density ($\sim$ 500 TOPS/$W/mm^2$). 2-D RAP and SAP show 20\% and 30\% increase in the peak performance compared to 1-D versions while keeping the same power density. This figure depicts an intra-class comparison against the performance of the state-of-the-art commercial digital accelerators such as Goya\cite{medina2020habana}, Google TPUv4\cite{wang_selvan}, GraphCore C2\cite{lacey}, Groq\cite{gwennap2020groq}, Nvidia A100 \cite{campa_kawalek_vo_bessoudo_2021}, and Tenstorrent \cite{Tenstorrent}. Such digital accelerators have around 1 Peta OPS/$W/mm^2$ with power density less than 0.5 $W/mm^2$. 
In addition to an inner-class IMC comparison is shown against resistive neural accelerator such as PUMA \cite{ankit2019puma}, Newton \cite{nag2018newton} and ISAAC \cite{shafiee2016isaac}. Such accelerators utilize resistive crossbar arrays for accelerating vector-matrix multiplication and are dedicated to deep neural networks acceleration. Such resistive accelerators have less than 100 TOPS/$W/mm^2$ for ISAAC and PUMA and 350 TOPS/$W/mm^2$ for Newton. APs outperform the inner-class IMC accelerators in terms of peak performance at the expense of higher power density. It is worth mentioning that the power density of RAP is around 5.5 $W/mm^2$ which is not acceptable. Most of these accelerators are mainly targeted for DNN models acceleration, unlike APs that can be placed in between general-purpose accelerators, such as CPUs, and the domain-specific accelerators such as resistive neural accelerators, TPU, and others. 

\subsection{Promising Applications}
APs have been explored for many applications such as matrix multiplication \cite{yavits2014sparse,neggaz2018rapid}, fast Fourier transform (FFT) \cite{yantir2019ultra}, discrete Fourier transform (D{F}T) and video application\cite{balam2006associative}, DNA sequence alignment \cite{kaplan2017resistive}, stencil applications \cite{yantir2020efficient}, convolution operation \cite{yantir2021imca}, solution of path problems\cite{nepomniaschaya1997solution},
optimum branchings\cite{nepomniaschaya2001efficient}, databases applications \cite{schuster1979rap} and computer vision \cite{krikelis1991computer}. The old applications need to be revisited and re-evaluated under the new AP design approaches and technologies besides exploring new applications that could benefit from the AP. Here we shed the light on two applications, one to be revisited and another to be explored.  

\subsubsection{\textbf{Speech-to-Speech Translation}}

One of the interesting applications for the associative processors is the real-time speech-to-speech translation (S2ST) which was and still is considered one of the grand challenges of Artificial Intelligence \cite{higuchi1994ixm2}. Such application requires many computations that have to be done in real-time and a massive capacity for handling vocabulary and the corpus. IXM is an AP designed targeting S2ST where 'IX' stands for semantic memory system in Japanese and 'M' for machine. Hence, APs could be a way to enable real-time S2ST, but, IXM2 needs to be revisited under the recent advances in natural language processing \cite{min2021recent}. 

\subsubsection{\textbf{Towards Decimal Computations}}
The associative memory (i.e., CAM) can be extended to store multi-valued data instead of binary data enabling high-density memories thanks to multi-valued devices \cite{wang2020resistive} and analog CAM cell\cite{li2020analog}. Such technology would require revisiting the binary computing paradigm towards multi-valued logic computations. So, new algorithms and architectures need to be developed to support multi-valued logic either through generalizing the existing algorithms such as in \cite{hout2021memory} or inventing new ones. In \cite{hout2021memory}, we generalized and optimized the LUT generation to support multi-valued logic. And we evaluated the ternary AP against binary AP showing a 12.25\% and 6.2\% reduction in the energy and the area, respectively.

\section{Challenges and Possible Solutions}
\label{Sec:limitations}
Due to the nature of the AP hardware, AP suffers from many limitations and challenges. We discuss some of them in this section. 

\textbf{\textit{Instruction Branching:}}
In theory, APs can execute any benchmark that traditional processors execute. However, APs achieve good performance and provide benefit for the problems that fit for SIMD architectures as discussed in Section \ref{Sec:AP_apps}. The architecture could limit the performance gain if the application includes too many branches and more sequential sections than parallel ones. In any case, the architecture can handle the branching through the predication, the technique used in GPU architectures \cite{GPUbranching}.

\textbf{\textit{Extensive Writes and Limited Endurance Cycles:}}
Since most AP operations are done inside the memory, extensive write cycles are needed which highly increase the total energy consumption to perform the operation and run out the CAM cell very quickly especially with memristor devices. For instance, the number of writes for 64-bit in-place addition is around 96 writes on average \cite{hout2021memory} out of $\sim 640$ cycles.  Hence, the average lifetime of a cell is $\sim 66$ seconds for $10^{10}$ endurance cycles at a 1 GHz clock. For practical reasons, a computer needs to run for 10 years, hence the CAM cell endurance should be greater than $10^{16}$ endurance cycles. Consequently, the most mature technologies per Table \ref{tab:CAMCell} are SRAM-based CAMs.      
Selective write techniques are highly desired for instance the order of passes can be changed \cite{hout2021memory}. 

\textbf{\textit{Handling Large Amount of Data Movement:}}
Another major limitation is data movement cost. If an application requires computation between the rows, they must be reorganized since AP can perform computation between the values on a row. However, still AP can perform very well in such applications if data-level parallelism dominates as in FFT\cite{yantir2019ultra}. While memory to AP or AP to AP data movement can be addressed and optimized while designing the data flow and the parallelism of the workload.

\section{Conclusion and Future Directions}
\label{Sec:future_Dir}

In-memory AP architectures are very promising as an accelerator for the workloads that have SIMD nature which is very everywhere nowadays. APs can achieve an order of magnitude speedup compared to the existing CPUs \cite{yantir2018two} and show comparable performance to the recent hardware accelerators.   Besides solving the challenges and limitations of the APs, discussed in Section \ref{Sec:limitations},  future directions for having an efficient AP are summarized in the following points.

\textbf{\textit{Multi-bit CAM Cell:}}
Recent works show the ability to fabricate multi-bit cells enabling much denser associative memories \cite{li2020analog,yin2020fecam,bazzi2022efficient}. Hence, finding ways to handle multi-bit devices and extending the functionality to multi-valued logic is of interest.

\textbf{\textit{AP Compiler:}}
To further explore and evaluate the functionality and the performance of various applications, an AP compiler is highly desirable to map different tasks into SIMD instructions to be run and evaluated on the AP. Such a compiler would help to have a realistic and fair assessment against other processors accelerators \cite{smagulova2021resistive}. An early work on an AP compiler is discussed in \cite{yantir2022hardware} where depth-first search (DFS)-based scheduler was introduced for energy and area optimization.

\textbf{\textit{AP Integration with Computing Platforms:}}
APs are with limited capabilities to SIMD only as a standalone processor, as discussed in Section \ref{Sec:limitations}. Such limitation could prevent deploying APs in some situations such as on edge acceleration which requires an external host for configuration and data input management. Hence, AP capabilities can be further extended when the AP is used as a coprocessor where the host processor is used to perform the non-SIMD tasks and data management \cite{herrmann1992dynamic}. Open source processors, such as RISC-V, could be integrated with the AP, where the host processor can be used also as a controller to AP. Another direction to further enhance the AP is to scale the AP architecture with multiple AP cores to support different computing operations such as multiple-instruction multiple data (MIMD) and revisit the old architecture such as associative string processors \cite{krikelis1991computer}.



\end{document}